# HyBIS: Windows Guest Protection through Advanced Memory Introspection

Roberto Di Pietro, Federico Franzoni, Flavio Lombardi

**Abstract**—Effectively protecting the Windows OS is a challenging task, since most implementation details are not publicly known. Windows has always been the main target of malwares that have exploited numerous bugs and vulnerabilities. Recent trusted boot and additional integrity checks have rendered the Windows OS less vulnerable to kernel-level rootkits. Nevertheless, guest Windows Virtual Machines are becoming an increasingly interesting attack target. In this work we introduce and analyze a novel *Hypervisor-Based Introspection System* (HyBIS) we developed for protecting Windows OSes from malware and rootkits. The HyBIS architecture is motivated and detailed, while targeted experimental results show its effectiveness. Comparison with related work highlights main HyBIS advantages such as: effective semantic introspection, support for 64-bit architectures and for latest Windows (8.x and 10), advanced malware disabling capabilities. We believe the research effort reported here will pave the way to further advances in the security of Windows OSes.

## 1 INTRODUCTION

Securing the Windows OS is a very challenging task, given its complexity and also given that its internals are not publicly known. Over time, a large set of malwares have targeted vulnerabilities in Windows OSes and services. Due to the very large installed base of Windows OSes, there is a great amount of new malware produced every year, which implements advanced methods for detection avoidance. The problem is particularly interesting for recent Windows versions, which have not yet been fully analyzed/investigated by the research community.

Among the different kinds of malware, rootkits represent the most complex and dangerous threats. In fact, rootkits can alter the system's perception of itself, and conceal malicious activities over a large period of time (i.e. APTs

• Roberto Di Pietro is with Bell Labs, Cyber Security Research, 91620 Nozay, Paris, France. He is also with Maths Dept. Univ. of Padua, Italy.
E-mail: roberto.di_pietro@alcatel-lucent.com (R. Di Pietro)
Federico Franzoni holds a M.Sc. in Computer Science from Sapienza University of Rome, Italy.
E-mail: fed.franzoni@gmail.com
Flavio Lombardi is with IAC-CNR, via dei Taurini 19, 00185, Rome, Italy.
E-mail: flavio.lombardi@cnr.it (F. Lombardi).

[1]). In particular, modern rootkits can directly manipulate memory structures to further enhance their stealthiness. As such, security tools can hardly detect them and are usually unable react to the infection. For this reason, rootkit detection is a vital task for protecting Windows and it is then fundamental to make it as effective as possible.

### 1.1 Motivation

Current monitoring approaches cannot provide an adequate level of protection against rootkits targeting Windows OSes. In fact, most present solutions operate at the same level as rootkits do[2], [3]. By tampering with the functions leveraged by security tools, rootkits are able to evade detection from within the OS. Hence, anti-rootkit tools working at the OS level cannot be trusted in case of rootkit infection. When the OS is running in a virtual machine, however, this problem can be addressed in a different way. Such a scenario, in fact, allows an external observation of the OS, from a more trustworthy and isolated environment. This capability is provided by the hypervisor, which can directly access VM components without leveraging OS functions.

Such a capability enables the adoption of virtual machine introspection (VMI) [4], [5],

which consists of inferring the guest OS semantics from the analysis of the status of VM components. VMI provides a valuable tool to counter rootkits since they can hardly conceal their presence to an monitoring system not dependent on OS functions.

On the one hand, VMI on Windows guest is however hard in practice as it requires some specific OS information to in order to make sense out of raw machine data[5]. This is one of the challenges of our present work, and it is also one of the main contribution of this work. On the other hand, VMI can be supported by the use of the forensic memory analysis (FMA), which provides the means for extracting OS information from raw memory data. In fact, as stated above, modern rootkits manipulate memory to avoid detection and can thus be identified by inspecting the same memory contents[6]. This is a clear advantage over rootkits and allows the implementation of more reliable security systems.

Moreover, once the infection has been identified, the hypervisor also allows an effective reaction. In fact, by leveraging unfettered full access to physical resources, a security tool can directly manipulate the VM and stop rootkit activities.

All these features, render the hypervisor a very attractive place where to implement security functionalities. In this work, we will leverage advanced VMI and FMA to help securing Windows OSes in virtualized environments.

### 1.2 Contribution

This work introduces and discusses a novel effective approach for countering rootkits on a Windows OS running in a VM. The implemented security monitor is external to the target machine, similarly to some recent literature[7], [8], [9]. By leveraging VMI and current FMA tools, we developed a novel *Hypervisor-Based Introspection System* (HyBIS) for protecting a Windows OS from stealth malware, in particular from rootkits.

The proposed system extends the hypervisor to monitor the state of the running machine, to detect rootkits, and to react to the discovered anomalies. The monitoring functionality leverages VMI techniques to infer the guest OS status. In order to detect rootkits, guest memory is scanned for kernel objects which may have been hidden. Such a scan is performed on memory dump files by means of FMA techniques and tools. Although such tools are typically used for offline analyses, the proposed system utilises them in a live way, during the system execution. To this purpose, HyBIS provides a novel dumping system which allows improving the performance of the memory acquisition task. Furthermore, a novel reaction approach is implemented that makes use of the hypervisor to manipulate memory contents while the virtual machine is running. This capability is leveraged to prevent the execution of detected rootkit processes. HyBIS allows detecting and reacting to rootkits effectively on Windows 8.1 and Windows 10 OSes. HyBIS successfully proves that the combination of VMI and FMA provides a valuable tool for countering rootkits on Windows OSes.

## 2 HYBIS: AN HYPERVISOR-BASED INTROSPECTION SYSTEM FOR WINDOWS

This section describes HyBIS, our solution combining FMA techniques with the VMI approach.

As mentioned above, the main goal of our work is to improve Windows security in virtualized environments. In particular, our research focuses on protecting such OS from rootkits.

In order to protect Windows from modern rootkits, we mostly focused our studies on the RAM component. Memory, in fact, stores both code and data and is involved in almost every operation performed on the machine during the OS execution. Thus, RAM can be considered the most complete source of information about the status of a running OS at a specific time.

For such a reason, Modern rootkits use to manipulate memory to conceal their activities and resources. Nonetheless they reside in RAM while running, thus giving the opportunity to detect their presence. Hence, memory is the best place where to look for inferring the current status of the target machine. FMA enables performing such a task in an effective and convenient way.

### 2.1 Our Approach

The basic idea we followed for the development of our security system was that the hypervisor can do more than what it is intended for. The chosen design approach was then to augment the hypervisor capabilities by means of introspection techniques. We extended the hypervisor by introducing the following functionalities:

- *Monitoring*: the hypervisor is enabled to to monitor the machine state in order to realize if something anomalous is happening;
- *Analysis*: the hypervisor is enabled to to analyze the state of the guest OS in order to detect the presence of rootkits;
- *Reaction*: the hypervisor is enabled to to react when a rootkit is detected and block its activities.

The above functionalities leverage internal hypervisor functions as well as external libraries and tools. The internal functions provide direct access to virtual machine hardware components. In particular, they allow monitoring the VM CPU and physical memory. By checking the CPU state and reading the memory contents, it is possible to implement a transparent monitoring function. Furthermore, the write access to memory can give the ability to perform changes into a running VM.

By making use of external tools and libraries, the hypervisor can be given even more capabilities. For instance, by integrating memory forensic functions, it is possible to implement advanced analysis techniques, which may allow detecting the presence of rootkits into the system.

**Monitoring: Checking The System State:** the virtual machine state can be analyzed by means of VMI. As stated before, the hypervisor has the ability to access virtual hardware resources directly, allowing the monitoring of all the VM components. In particular, we chose to monitor CPU and memory as they are core components of the machine.

The CPU state changes can be easily monitored by using internal hypervisor functions. Such functions allow, for instance, checking the current operating mode, or inspecting registers.

VM memory contents changes can be monitored by means of differential dumps. With this approach, memory dumps are periodically generated to check if a particular area has been modified. In order to implement such a functionality, an initial memory snapshot must be taken at a specific time. Such a snapshot can then be used as basis for the comparison with the following checkpoints.

**Analysis: Detecting Rootkits:** as previously explained, modern rootkits are able to manipulate memory objects at runtime to conceal their activities. Hence, in order to detect rootkits, the analysis functionality should focus on the memory contents. This kind of analysis requires advanced forensic techniques to discover an infection. A convenient approach would be then to make use of functions from an external forensic tool or library.

Memory forensic analyses are commonly based on memory dumps. Hence, a memory acquisition functionality is required to make use of the forensic tools. Fortunately, most hypervisors implement their own dumping facility which can be used to acquire guest memory. Such a facility can be easily expanded/adapted to improve the acquisition process and realize the above-mentioned differential dumping functionality.

**Reaction: Countering Rootkits:** in order to react when a rootkit is detected, some kind of action must be taken to prevent it to perform further activities. As in previous cases, the guest memory can be used to implement the reaction functionality.

Since rootkits, even if concealed, reside in memory, it would be a good approach to counter them into the same place. Once again, hypervisor functions can be helpful: by writing into guest memory it could be possible to delete the detected rootkit from memory, or to block its execution.

## 3 HYBIS FUNCTIONALITIES

We set up to implement four high-level functionalities in HyBIS:

- Automatic boot dump generation;
- Smart differential dumping;
- Detection of hidden rootkit processes;
- Blocking of hidden rootkit processes;

**Automatic Boot Dump Generation:** as explained above, the monitoring functionality should allow deciding when an analysis operation would be appropriate.

Since the analysis functionality operates over memory dumps, we decided to automatically generate a dump on the basis of some hardware event. In particular, we chose to monitor the VM during the boot phase in order to produce a dump at the very beginning of the Windows loading.

This choice has a twofold reason. Firstly, it aims at determining the first feasible moment for analyzing a memory dump with a forensic tool; in fact, these tools need the kernel to be loaded in order to work. Secondly, such a preliminary dump can be used as starting point for a following monitoring of the memory; in fact, after the kernel has loaded, most of the system areas remain fixed during the rest of OS execution.

Hence, this function should allow HyBIS to automatically generate a memory dump as soon as the Windows kernel process starts.

This objective has been chosen to demonstrate how, by means of introspection, the VM state monitoring can be effectively used for determining meaningful moments of the OS execution.

**Smart Differential Dumping:** besides the CPU, the monitoring functionality can check specific virtual machine memory areas for changes, in order to decide if an analysis operation is needed.

As stated above, a differential approach can be taken to perform this kind of monitoring. However, memory acquisition can be a very onerous task to perform, especially when it has to be repeated over time. So, it is important to do such an operation efficiently in order to not compromise the guest system performance. Since only some memory ranges need to be checked, there is no need to dump the whole memory at every checkpoint. Instead, it should be enough to acquire only the ranges we are interested in.

This function should allow HyBIS to update a previously created dump by acquiring selective ranges and overwriting them into the corresponding areas. Previous contents of such ranges should be backed up in separate files in order to allow the comparison between different checkpoints. With such an approach, it can be said that HyBIS uses "dynamic" dumps.

Dynamic dumps can also be used to improve the acquisition process necessary for the forensic analyses. In fact, since such analyses usually involve only some ranges of the whole dump, it is possible to use the update mechanisms described above, to perform the analyses of different checkpoints without needing to create multiple dumps of the whole memory.

This objective has been chosen to demonstrate how monitoring the VM memory can be both effective and efficient.

**Detection of hidden rootkit processes:** it is well-known that rootkits try to hide their processes to avoid detection.

This is effectively obtained by implementing the DKOM technique.In particular, a rootkit who wants to hide a process could remove the corresponding object from the active process list. In fact, Windows uses two list of processes: one for the scheduling, and one for tracking. A process whose object is removed from the tracking list, will be invisible while still active.

As such, hidden processes can be detected by means of a cross-view analysis. More specifically, this can be done by scanning memory for process objects, and comparing results with the active process list. If a scanned process is not present in such a list it is likely to be a hidden rootkit process. This function should allow HyBIS to detect hidden processes by creating a memory dump and scanning it for concealed process objects.

This objective has been chosen to demonstrate how FMA can help in detecting rootkits on running guest OSes.

**Blocking of hidden rootkit processes:** once a hidden process has been detected, it should be blocked to prevent it from keeping performing malicious activities. This action will not clean the infection but it could be a first step to defeat the rootkit.

A good idea for blocking a hidden process would be to exclude it from scheduling, thus preventing its execution. This can be done using the DKOM technique in a similar way as that used by the rootkits. More specifically, we can block an hidden process by removing the corresponding object from the scheduling list. This function should allow HyBIS to manipulate the VM memory in order to prevent the rootkit process to be executed.

This objective has been chosen to demonstrate how the hypervisor capabilities allow an effective reaction to a rootkit infection by means of memory manipulation.

## 4 DESIGN AND ARCHITECTURE

This section shows how HyBIS was designed to extend the hypervisor capabilities for securing a guest Windows OS from the rootkit threat. First we show the overall HyBIS operation from a high-level point of view. Next, we describe the HyBIS architecture. Finally, we describe and motivate the technologies chosen to build up the first HyBIS prototype, and the most relevant implementation details.

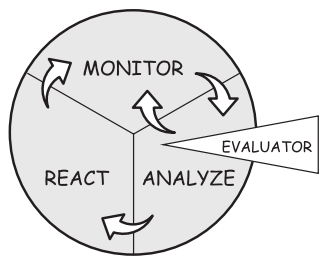

Figure 1. HyBIS Closed-Loop Design

The high-level overview of the HyBIS operating mode is depicted in Figure 1. The new functionalities are designed to work as a closed loop control system. The monitoring phase extracts information from a running machine and intercepts events which could reveal the presence of rootkits. The analysis phase examines the system in order to evaluate if a rootkit infection occurred. In such a case, it triggers the reaction phase, otherwise it returns to the monitoring phase. The reaction phase tries to remove the infection or block the rootkit for preventing further malicious activities.

The monitoring and reaction functionalities leverage the introspection and control capabilities of the hypervisor. The analysis functionality is based on forensic functions provided by external tools but needs some additional intervention to interpret the results and taking further actions. Since the complexity of such a task, some kind of intelligence is needed to take decisions. This is represented by the *evaluator*, which is an external component that can be inserted into the analysis phase. The evaluator functionality can be performed by a human examiner as well as an external plugin which implements advanced AI techniques[10], such as Machine Learning[11], Expert Systems [12], Human Expertise, and so on.

### 4.1 Architecture

The HyBIS architecture is shown in Figure 2.

As can be seen, on the guest side, there is a Windows OS running on a virtual machine (VM). On the host side, there is the hypervisor that controls the VM, which incorporates the HyBIS component. The HyBIS component extends the hypervisor with the new security functionalities. Such functionalities are implemented by three components, described below. Outside the hypervisor, on the same side, there is the memory forensic tool, which is used by HyBIS to provide advanced analysis capabilities.

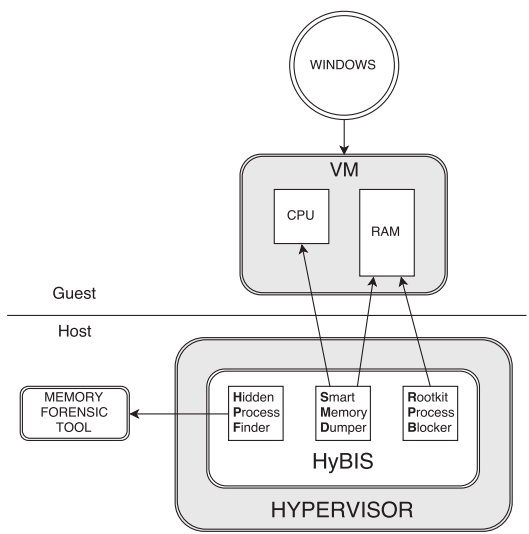

Figure 2. HyBIS Architecture

**HyBIS Components:** HyBIS includes three components:

- *Smart Memory Dumper* (SMD): this component allows creating the dynamic dumps described in the previous section; it leverages the hypervisor to read the VM's RAM contents and create or update dump files on the host disk.
- *Hidden Process Finder* (HPF): this component allows detecting hidden processes running in the guest Windows OS; it leverages the external memory forensic tool to perform analyses on the dump files created by SMD.
- *Rootkit Process Blocker* (RPB): this component allows blocking a detected rootkit process on the guest OS, by preventing it from being scheduled for execution; it receives from HPF the information on the detected process and leverages the hypervisor for manipulating the VM's RAM.

All these components operate while the guest OS is running, without interrupting/suspending its execution.

## 4.2 Technology Details

In this section we discuss the technologies selected for the development of the first HyBIS prototype. In particular, the target Windows version, the hypervisor and the forensic tool have been chosen due to their effectiveness and wide deployment base.

**The Target OS:** most of the latest security-related work still focuses on Windows XP or Windows 7 OSes. However, Windows 8 introduced some internal changes (such as [13]) and security mechanisms (see [14]) which partially invalidate previous results. For instance, the removal of the `KiFastSystemCall` function makes all rootkit techniques based on this function unusable[15]. Furthermore, the latest Windows 10 OS appears to keep such changes, rendering previous work yet more obsolete.

At the time we started the development of HyBIS, Windows 10 was only available in its Technical Preview release. Hence, we selected Windows 8.1 (which is much more widespread then Windows 8) as the target of our experiments.

We initially decided to focus our tests on the 32-bit version since it is more efficient when performing extensive memory-related experiments. Furthermore, the 64-bit version which implements more advanced security mechanisms, would have limited our malware testbed. As such, it will be the target of future work.

**The Hypervisor:** most of the recent projects targeting Windows as guest OS, involve the qemu-kvm [16] or the Xen hypervisors [17] (e.g. [18], [19]). Although these ones represent valid tools, we decided to make use of the VirtualBox[20] hypervisor. In fact, VirtualBox has two main advantages over qemu-kvm and Xen: firstly, it fully supports all Windows versions, including the latest Windows 10; secondly, it includes various VM-debugging functionalities, that allow controlling and manipulating VM components[21]. Such funcionalities can be very useful when implementating advanced introspection techniques.

For the HyBIS prototype implementation the latest VirtualBox 5.0 version has been used.

**The Memory Forensic Tool:** among the available FMA tools, Volatility[22] is certainly the most widespread. It has a vast number of functionalities and it can count on a very active community. Nonetheless, it does not fully support all Windows kernel versions. In addition,

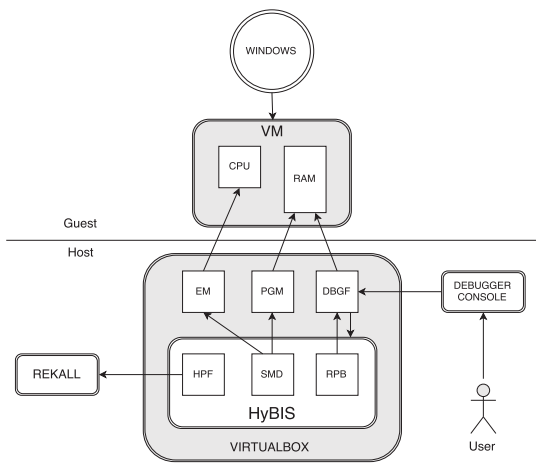

Figure 3. The HyBIS prototype leverages the VirtualBox's Execution Manager (EM), Page Manager (PGM), and Debug Facility (DBGF) components

its performance on memory analysis is quite low for our real-time usage requirement.

A derived project, named Rekall[23], overcomes these limits, while maintaining main Volatility features and advantages. Its novel kernel profiling system, enables Rekall to automatically upgrade its compatibility with new Windows versions[24]. Furthermore, thanks to some improved memory-scanning functions, it shows better analysis performances. This renders Rekall both more efficient and effective than Volatility. Moreover, it can be integrated as part of other software as a library. Finally, since Rekall is implemented in the Python language, it can be easily installed into a variety of host OSes.

## 5 IMPLEMENTATION DETAILS

We will now describe how the proposed architecture was implemented by making use of the previosly-mentioned technologies.

The HyBIS implementation details are depicted in Figure 3. As before, on the guest side we have a Windows OS, running in a VM. On the host side, we have the VirtualBox hypervisor, with its internal components: the Execution Manager (EM), the Page Manager (PGM), and the Debug Facility (DBGF). The HyBIS component is implemented as a new VirtualBox component, interacting with the other ones to perform its task. In particular, the SMD component leverages PGM for the memory acquisition, and EM for the automatic boot dump generation. The RPB component uses DBGF to manipulate guest memory. The HPF component makes use of the external Rekall component to perform advanced memory analyses over memory dumps.

The interaction with the HyBIS components is provided through a set of new commands on the VirtualBox integrated Debug Console. These commands will be described later in this section.

### 5.1 Extending VirtualBox

The following VirtualBox components have been involved for the HyBIS implementation:
- Virtual Machine Monitor (`VMM`): this is the core hypervisor.
- Execution Manager (`EM`): it controls the execution of guest code.
- Page Manager (`PGM`): it controls guest memory paging.
- Debug Facility (`DBGF`): it provides a built-in debugger for the VM.

Most of the VM-related components are implemented in the VMM section. Therefore, this is the most suitable place where to insert the new HyBIS component.

The EM component is leveraged to monitor the CPU operation mode in order to automatically generate the boot dump. The PGM provides all guest memory management functions and is used to implement the acquisition functionalities. The DBGF provides a lot of useful debugging functions, which are used for implementing the reaction functionality. Furthermore, it provides the console devoted to the interaction with the HyBIS component.

### 5.2 Integrating Rekall

Since Rekall is written in Python, an interpreter must be present on the host system.

For incorporating Rekall, it has been necessary to import the Python C++ library into the VirtualBox source code. The Rekall functionalities, instead, can be used by means of the provided API library, as mentioned in the previous section. In order to make use of the forensic analysis functions, a suitable session has to be set up. A Rekall session represents a specific combination of a dump file and a selected profile.

When starting a session, if a valid profile for the current kernel version is found, every compatible Rekall plugin automatically becomes available to be used for a forensic analysis.

### 5.3 Using HyBIS

In order to enable the HyBIS funcionalities, the VirtualBox sources must be patched and recompiled. To this purpose, a generic HyBIS patch has been created, which is easily applicable the almost every recent VirtualBox version. Furthermore, it is necessary to run VirtualBox with the debug option (`--dbg`) to enable the interactive console.

As previously mentioned, HyBIS implements its own debugger commands to allow using the new functionalities. The main commands are:
- `.dumpmem`: generates a raw dump file;
- `.dumprangediff`: updates a specific memory area in a previous dump, saving overwritten data in separate files;
- `.setbootdump`: sets an automatic memory dump at startup (needs reboot);
- `.pslist`: scans guest memory for processes and prints a compared view table;
- `.psblock`: receives the address of a process object and removes it from the scheduling list;
- `.startsess` and `.stopsess`: start/stop a new analysis session;

Two kinds of interaction are supported: (1) the *standalone* mode and (2) the *session* mode. The *session* mode creates a dump file and allows following actions to be taken on it. This allows performing multiple analyses over the same dump. In the *standalone* mode, every submitted command creates a temporary dump on which the command will be applied. This mode leaves the currently open session untouched, enabling the comparison of the results of a test with a previous checkpoint.

## 6 EXPERIMENTAL RESULTS

A large number of experiments have been performed for testing the implemented prototype. Both effectiveness and functionality of each HyBIS component described in Section 3 has been tested. In this section we show the testbed used for the experiments and the preliminary results obtained.

**Testbed:** All experiments were conducted on a VM with 1 GB of memory and a single processor with Intel VT-x, EPT, and PAE enabled. As stated above, the target OS was Windows 8.1 32-bit. The OS was infected with different rootkit specimens, chosen among the most widespread and dangerous, like ZeuS[25], or ZeroAccess[26].

**Boot Dump Generation:** In order to prove the effectiveness of this function, the generated dumps has been analyzed with Rekall. In particular, after scanning the dump for active processes, the only `System` process has been found. This proves that the dump generation occurs at the beginning of the Windows kernel execution.

In addition, for testing purposes, some dumps have been automatically generated just before the "correct" time. Both Rekall and Volatility, failed to analyze such dumps in any way. This is due to the fact that such tools rely on specific memory objects created by the kernel.

As such, we can state this function generates a dump at the very first suitable moment for being analyzed by these FMA tools. Moreover, we believe this functionality can be helpful in detecting rootkits which load very early during the Windows boot phase.

**Dynamic Dumping:** This function has been tested by updating a restricted area of a dump. In particular, we identified a range of addresses, which we showed to always contain all process objects. Such a range has size 250 MB, that is a quarter of the original 1 GB size.

With the purpose of proving the usefulness of this functionality, we started a new process in the guest, before the updating the dump. Then we scanned the updated dump for active processes by using Rekall. As expected, the resulting process list also contained the new process.

This demonstrates that such a dynamic memory acquisition does not prevent Rekall to properly work with respect to specific operations. This enables a new form of memory monitoring, which is not limited to single page changes, but involves larger areas. In fact, by leveraging FMA, it is possible to check such areas for more "high-level" changes, such as the presence a new process.

**Detection and Blocking of Hidden Processes:** By comparing the process list returned by the new `.pslist` command of the VirtualBox debugger, to that of the guest Task Manager, we have been able to detect various hidden processes created by rootkits.

The Rekall plugin used to implement the command probes memory using various methods, including a full scan for process objects. Since the Windows kernel does not immediately delete objects after a process terminates, it is possible to have a lot of false positives in the results. Fortunately, these are easily recognized in that they only results from the full scan, while being absent from all system lists.

After identifying the hidden process, it was possible to block it by means of the `.psblock` command. This removes the hidden process object from the system scheduling list (and other system lists), thus preventing its execution. In order to prove the effectiveness of this action, we analyzed again the active processes and verified the malicious process was not in the active process lists anymore.

**Considerations on Windows 8.1:** During the experiments, we tried to install a huge number of rootkit specimens on the guest Windows OS, but only a few of them successfully infected the system. Many of them fail to run, while others achieve to run, but even when executed with privileges, they fail to tamper with the target resources. In particular they often failed to create hidden persisten processes.

Conversely, when testing the same specimens against Windows 7, they mostly succeed in corrupting the OS. This fact may be due to more effective protection of critical resources by the Windows 8.1 kernel, or it can be caused by a lack of rootkits targeting such a kernel version.

## 7 DISCUSSION

HyBIS embodies several novel features, for both its architecture and its implementation.

### 7.1 Architecture

Technically speaking, HyBIS is an hypervisor-based IDS which leverages Virtual Machine Introspection (VMI) to monitor a virtualized environment and exploits Forensic Memory Analysis (FMA) to bridge the semantic gap.

With respect to previous solutions, HyBIS uses a novel approach, whose differences are explained in the following.

999

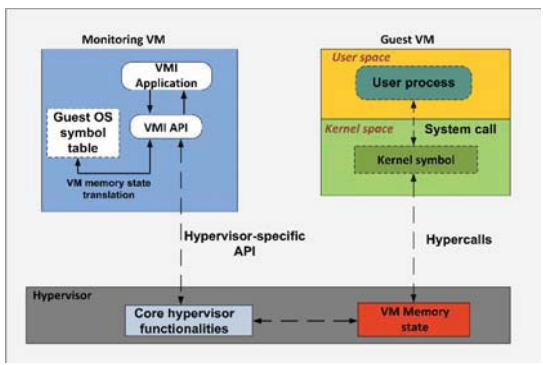

Figure 4. A Typical VMI Architecture[27]

**Use of VMI:** typical hypervisor based IDSs reside on the top of the hypervisor, in a Secure Virtual Machine (SVM) (see Figure 4).

Such an approach has the advantage of isolating the security functions so to avoid corruption. However, in such a case, the IDS requires an additional VM, and its capabilities are limited by the functionalities provided through the hypervisor APIs. HyBIS, instead, is integrated into the hypervisor, as shown in Figure 2. As such, it is able to exploit all the hypervisor capabilities in order to have a full control over the target VM.

An alternative VMI approach leverages advanced CPU features to interpose the security functions between the OS and the hardware[28], [29]. However, this kind of hypervisor does not have the control over all the machine components. Furthermore, this approach usually requires an in-guest agent to be installed into the OS kernel. Conversely, HyBIS does not require any addition to the guest OS kernel and does not rely on any CPU feature.

So, while current VMI-based solutions have limited possibilities to exploit the hypervisor capabilities, the HyBIS novel approach gives the ability to fully exploit them by integrating into the hypervisor. By working from below the VM, it has a full overview of the whole machine state, while is still isolated from the target machine, thus avoiding both OS-level and hardware-level attacks.

**Use of Memory Forensics:** FMA can be a valuable means for VMI-based systems to bridge the semantic gap. However, it has not yet been fully leveraged by current solutions. This is probably due to the slowness of the memory acquisition process. In fact, FMA tools usually operate on offline dump files, whose creation may take too long for a practical real-time usage. As such, memory acquisition, represents a critical step for the implementation of a real-time FMA-based solution. Furthermore, the soundness of the acquisition process is another serious concern[30].

Typically, memory is dumped by tools installed into the guest OS (see [31]). Such tools have the drawback of altering memory contents and to be vulnerable to OS-level attacks which may corrupt acquired data. Alternatively, memory acquisition can be hardware-based, using techniques, such as DMA, that allow bypassing the OS. Such techniques, however, can still be bypassed by hardware-level rootkits, such as in [32] [33] and [34], which are able to alter acquired data for hiding their presence.

By leveraging the hypervisor, HyBIS is able to acquire memory, without being vulnerable OS-level and hardware-level attacks. Also, its novel dynamic approach allows an efficient dump creation, thus enabling a live usage of the FMA tools.

## 7.2 Implementation

The chosen technologies for the HyBIS implementation was not only chosen for their useful features. All of them represent an element of novelty for security research.

First of all, the chosen Windows kernel version has been poorly explored in previous work. Most recent Windows-related papers still refer to Windows 7 as the subject of their studies, or as the target for their experiments. Actually we was not able to find any kernel-related research on Windows 8 and Windows 8.1. Instead, in our HyBIS implementation and testing Windows 8.1 was used. The latest Windows 10 has also been successfully tested against our prototype.

Leveraging the Virtualbox hypervisor represents another relevant contribution of this paper. Most security and virtualization studies involved other common hypervisors, like KVM or Xen. Instead, besides a few performance analysis, we was able to find only a single work involving VirtualBox for its implementation[35]. In the development of HyBIS, VirtualBox resulted as a great tool to be employed for our project. In fact, its features and functionalities have been very helpful for the exploration of VMI techniques.

A similar discussion can be made for the Rekall forensic tool. To the best of our knowledge, Rekall was rarely used as part of a security research project. The only online reference found, was in [36], where Rekall was used to analyze Windows profiling. This is probably due to its relatively recent introduction (2007). Most studies and researches involved Volatility instead, from which Rekall was derived. In fact, Volatility has the advantage of being more widespread and supported. However, as previously described, Rekall presents almost the same features but introduce novel features that drastically improves performance and usability. As such, we found it a valuable tool for memory forensic research.

## 8 Related Work

In this section we survey most relevant related work and stress main differences with respect to our solution.

Zhang et al.[37] leverage SMM, an advanced x86 execution mode, for detecting memory-based stealthy malware. Their SPECTRE framework can introspect a live operating system and supports both Windows and Linux OSes. However, this framework is vulnerable to hardware-based attacks, such as [34]. Furthermore, their work is limited to Windows XP SP3.

In [38], Hizver and Chiueh make use of Volatility for the analysis of VM execution states. Their RTKDMS system is able to perform real-time monitoring at the hypervisor level. Differently from HyBIS, such an architecture leverages an additional VM for the introspection analysis. Again, the experiments are limited to Windows XP. Furthermore, their system does not tackle the rootkit threat specifically, neither it explores any reaction possibility.

Deng et al. [18] propose the SPIDER archicture, a stealthy program instrumentation and debugging framework built upon hardware virtualization. SPIDER enables monitoring memory read/write at any address. Nonetheless, unlike HyBIS, it requires an in-guest agent which modifies the guest OS kernel.

In [19], Lengyel et al. describe DRAKVUF, a novel dynamic malware analysis system based on Xen, which improves hardware resources usage efficiency. DRAKVUF takes advantage of the hardware virtualization extensions to provide a transparent and scalable environment to enable in-depth analysis of malware samples. In this case, the target OS is Windows 7 SP1 in both 32- and 64-bit versions.

Zhang et al. [34] hacks CPU registers, dedicated to the MMIO mechanism, to conceal a memory portion used to store malicious code. In-guest software is unable to access the hidden memory portions since all their operations pass through the CPU. The proposed mechanism is also able to evade physical memory forensic though DMA. Since HyBIS operates at the hypervisor level, it does not rely on the guest machine and it is then able to access the whole physical memory without restrictions. Hence, HyBIS is not affected by this kind of rootkit. Furthermore, at the hypervisor level it is possible to monitor the CPU in order to detect possibly unsolicited register modification and then to prevent this type of hacks.

Harrison[7] suggests an approach that is somewhat similar to HyBIS. However it aims to be integrated into other IDS solutions and mostly focuses on the analysis phase. HyBIS, instead also explores the novel reaction capabilities given by the combination of VMI and FMA techniques.

## 9 Conclusion and Future Work

In this work we provide several contributions, shedding light on some security issues of the obscure Windows security computing field. The main contribution is the design of the novel HyBIS architecture, which successfully combines VMI and FMA to build up an anti-rootkit security system for Windows. VMI is used to examine the Windows status by means of hardware monitoring, while FMA is used to carve meaningful information from the raw memory data. Initial experimental activity was performed over most relevant malware specimens, allowing us to detect and block different hidden processes. Given the generality of the architectural complexity of the Windows kernel security field, the results reported in this paper—other than being interesting on their own—also pave the way for further research.